\documentclass[a4paper,11pt]{article}
\usepackage{pos}

\title{$H$ dibaryon away from the $SU(3)_f$ symmetric point\note{MITP-21-061}}
\author*[a]{M.\ Padmanath}
\author[b]{John Bulava}
\author[c,1]{Jeremy R.\ Green}
\author[d]{Andrew D.\ Hanlon}
\author[e]{Ben~H\"orz}
\author[f]{Parikshit Junnarkar}
\author[g]{Colin Morningstar}
\author[h]{Srijit Paul}
\author[a,c,i,h]{Hartmut Wittig}

\affiliation[a]{Helmholtz Institut Mainz, Staudingerweg 18, 55128 Mainz, Germany, \\
GSI Helmholtzzentrum für Schwerionenforschung, Darmstadt (Germany)}

\affiliation[b]{Deutsches Elektronen-Synchrotron (DESY)
Platanenallee 6, 15738 Zeuthen, Germany}

\affiliation[c]{Theoretical Physics Department, CERN, 1211 Geneva 23, Switzerland}
\note{Present address: School of Mathematics and Hamilton Mathematics Institute, Trinity College, Dublin 2, Ireland}

\affiliation[d]{Physics Department, Brookhaven National Laboratory, Upton, New York 11973, USA}

\affiliation[e]{Nuclear Science Division, Lawrence Berkeley National Laboratory, Berkeley, CA 94720, USA}

\affiliation[f]{Institut f\"ur Kernphysik, Technische Universit\"at Darmstadt, \\ Schlossgartenstrasse 2, D-64289 Darmstadt, Germany}

\affiliation[g]{Department of Physics, Carnegie Mellon University,  Pittsburgh, PA 15213, USA}

\affiliation[h]{Institut f\"ur Kernphysik, Johannes Gutenberg-Universit\"at Mainz, \\
Johann-Joachim-Becher-Weg 45, D 55128 Mainz, Germany}

\affiliation[i]{PRISMA$^+$ Cluster of Excellence, Johannes Gutenberg-Universit\"at Mainz, \\
Staudingerweg 9, 55128 Mainz, Germany}

\emailAdd{pmadanag@uni-mainz.de}

\abstract{We present the current status of our efforts in search of $H$ dibaryon on $N_f$=2+1 CLS ensembles away from the $SU(3)$ flavor symmetric point. Utilizing the distillation framework (also known as LapH) in its exact and stochastic forms, we calculate two-point correlation matrices using large bases of bi-local two-baryon interpolators to reliably determine the low-energy spectra. We report the low lying spectrum on several moving frames for multiple ensembles with different lattice spacing and physical volumes. The status of finite-volume analysis to extract the scattering amplitudes is also discussed.}

\FullConference{%
 The 38th International Symposium on Lattice Field Theory, LATTICE2021
  26th-30th July, 2021
  Zoom/Gather@Massachusetts Institute of Technology
}

\newcommand\bef{\begin{figure}}
\newcommand\eef[1]{\label{fg:#1}\end{figure}}
\newcommand\bec{\begin{center}}
\newcommand\eec{\end{center}}
\newcommand\besf{\begin{subfigure}}
\newcommand\eesf[1]{\label{sfg:#1}\end{subfigure}}
\newcommand\beq{\begin{equation}}
\newcommand\eeq[1]{\label{#1}\end{equation}}
\newcommand\beqa{\begin{eqnarray}}
\newcommand\eeqa[1]{\label{#1}\end{eqnarray}}
\newcommand\bet{\begin{table}}
\newcommand\eet[1]{\label{tb:#1}\end{table}}
\newcommand\best{\begin{subtable}}
\newcommand\eest[1]{\label{stb:#1}\end{subtable}}
\newcommand\betb{\begin{center}\begin{tabular}}
\newcommand\eetb{\end{tabular}\end{center}}
\newcommand\beit{\begin{itemize}}
\newcommand\eeit{\end{itemize}}

\newcommand{\cref}[1]{\bec  \textcolor{black}{\small #1}\eec}

\newcommand\x[1]{{\mathbf{#1}}}

\definecolor{DarkGreen}{rgb}{0.00,0.29,0.00}
\definecolor{DarkRedfooter}{rgb}{0.60,0.00,0.00}
\definecolor{DarkRed}{rgb}{0.60,0.00,0.00}
\definecolor{DarkRedtitle}{rgb}{0.85,0.00,0.00}

\def\prsp#1#2%
  {\mathop{}%
   \mathopen{\vphantom{#2}}^{#1}%
   \kern-\scriptspace%
   #2}

\begin{document}
\maketitle

\section{Introduction}

A variety of tetra- and pentaquark states ({\it e.g.} $P_c$, $T_{cs}$, $T_{cc}$) was discovered in recent years, raising the scientific interest 
in such systems. Even so, despite various experimental efforts, 
there are only two six quark systems (deuteron and $d^*$(2380)) that are established to date. 
The existence of a deeply bound $SU(3)$ flavor singlet dibaryon with scalar quantum numbers,
referred to as $H$ dibaryon, was conjectured in 1977 \cite{Jaffe:1976yi}. While there is no concrete
experimental evidence in this regard, an upper bound of $\sim7$ MeV on the binding energy for
such a state relative to the $\Lambda\Lambda$ threshold was reported based on the constraints from the
Nagara event \cite{Takahashi:2001nm}. A recent study of the $\Lambda\Lambda$ interactions in p-p and p-Pb
collisions also reports results compatible with the existence of a shallow bound state \cite{ALICE:2019eol}.
With higher statistics from future runs at the LHC, the scattering parameters are expected to get
constrained further. 

The first lattice QCD calculation addressing the existence of a bound $H$ dibaryon was performed in 1985
\cite{Mackenzie:1985vv}. Since then, there have been several lattice calculations to date. Apart from 
the calculations by the Mainz group, calculations with dynamical quarks were performed by only two 
groups: HALQCD \cite{Inoue:2010es} and NPLQCD \cite{NPLQCD:2010ocs,NPLQCD:2012mex}. The calculation by 
the HALQCD collaboration was performed along the $SU(3)$ flavor symmetric line with varying pion masses. 
A calculation by the NPLQCD collaboration with an 800 MeV pion mass along the $SU(3)$ flavor symmetric 
line finds twice the binding energy as extracted by HALQCD at approximately the same pion mass. The 
NPLQCD collaboration reported a calculation with broken $SU(3)$ flavor symmetry in the other work. 
A general observation from these calculations is that the estimates for the binding energy decrease 
with decreasing pion masses. However, a clear consensus on the existence of such a state in the physical
limit from lattice calculations has not been reached. 

Lattice results from the Mainz group using $N_f$=2 ensembles indicate the existence of a bound $H$ dibaryon 
at heavier than physical pion masses in an $SU(3)$ flavor symmetric and broken setup with a quenched 
strange quark \cite{Francis:2018qch}. Recent results from an extensive study using $N_f$=2+1 ensembles with
five different lattice spacings also point to the existence of a shallow bound state, with significant
cut-off dependence in the lattice estimates \cite{Green:2021qol}. These calculations utilize the
finite-volume quantization condition {\it \`a la} L\"uscher to extract the infinite-volume binding energy.
The results at the $SU(3)_f$ symmetric point were discussed in a separate talk \cite{Jeremy:2021mai}. In this talk, we present
the status of Mainz efforts on $H$-dibaryon spectroscopy away from the $SU(3)_f$ symmetric point.

\section{Methodology}\label{method}

\begin{figure}
 \centering
 \includegraphics[height=6cm,width=5cm]{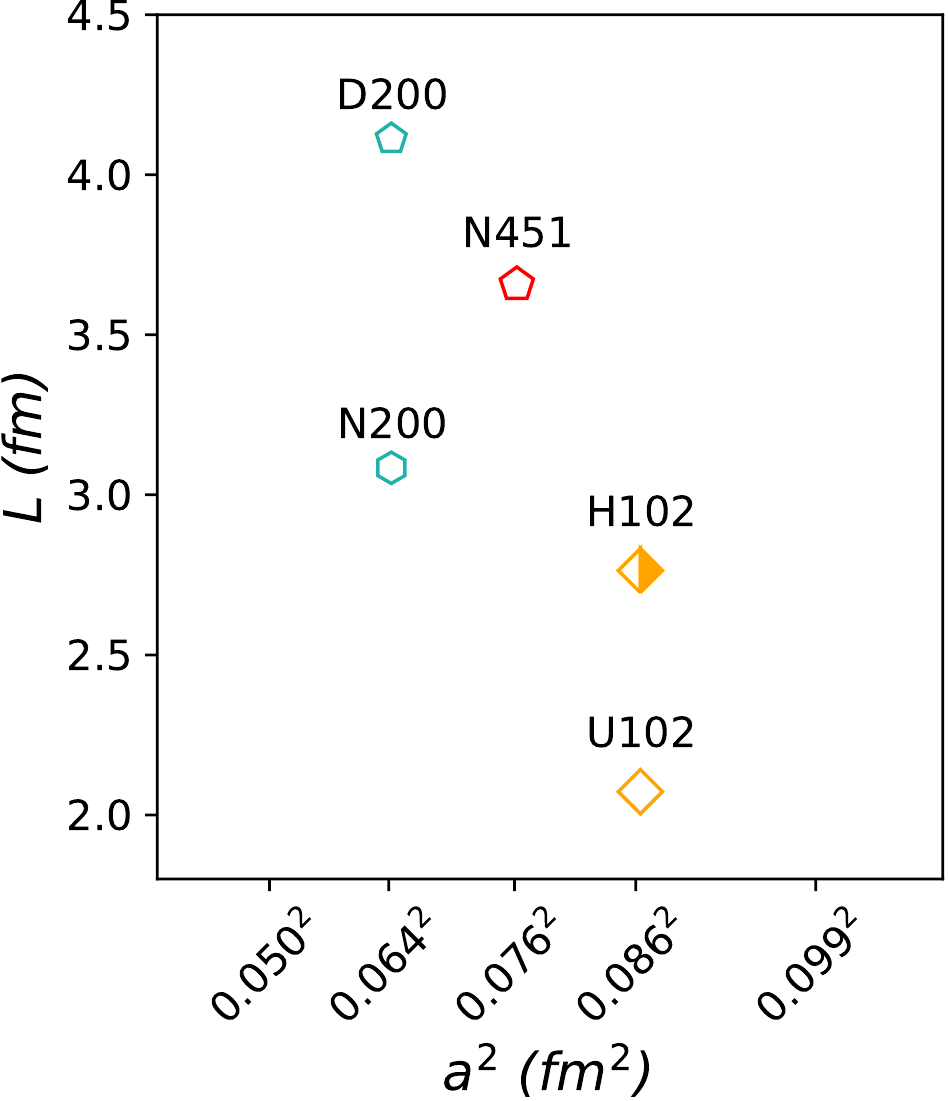}\hspace{1.5cm}
 \includegraphics[height=6cm,width=5cm]{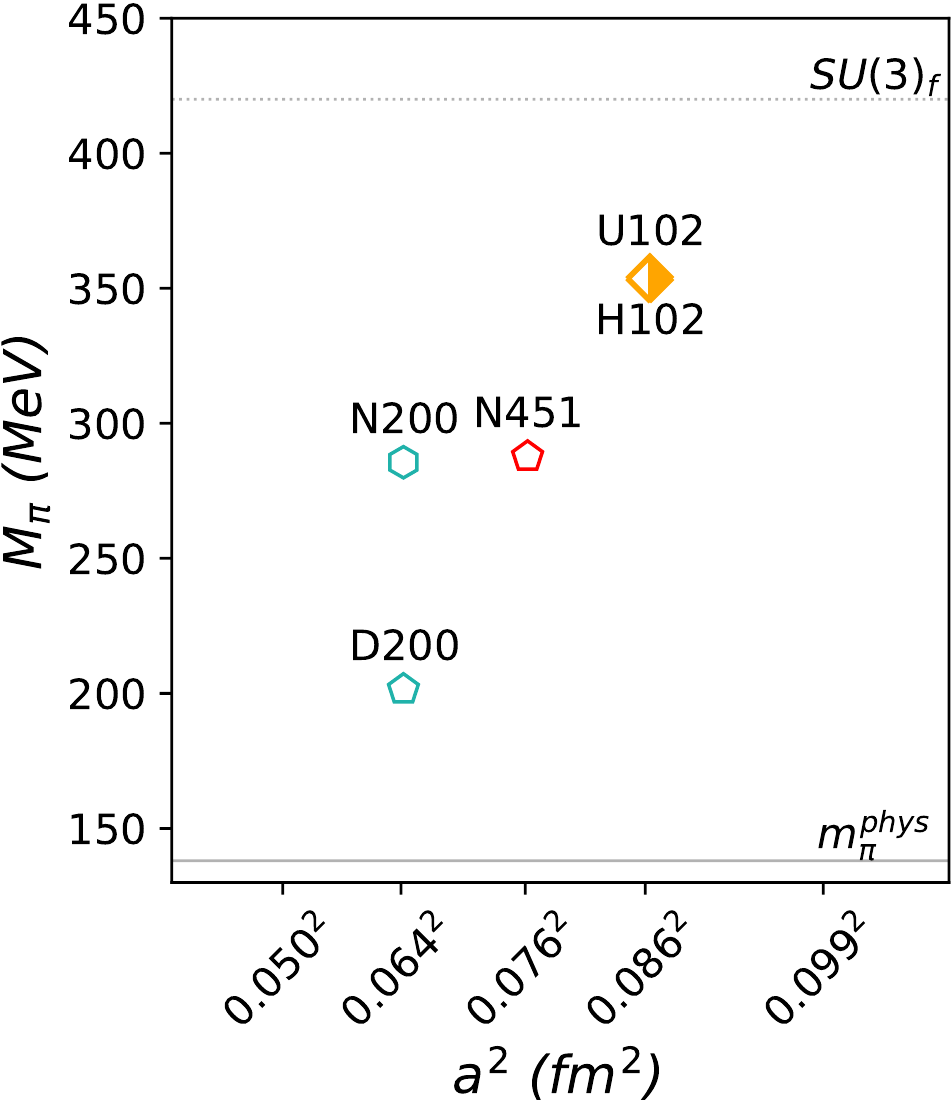}
 \caption{Left: Scatter plot of ensembles with y-axis referring to the physical lattice extension and 
 the x-axis gives the info on the lattice spacings. Right: The same set of ensembles with the y-axis 
 indicating the respective pion masses.}
 \label{ensmap}
\end{figure}

\textbf{Ensembles}:
We utilize the $N_f=2+1$ ensembles generated as a part of the Coordinated Lattice Simulations (CLS) 
effort. These ensembles have been generated with a nonperturbatively $\mathcal{O}(a)$ improved Wilson
fermion action and a tree-level $\mathcal{O}(a^2)$ improved L\"uscher-Weisz gauge action. All ensembles 
discussed in this talk lie on the $\textrm{Tr}(m)=2m_{u/d}+m_s =\textrm{const}$ trajectory that goes
through the physical point. The $SU(3)_f$ symmetric point on this trajectory is around $m_{\pi}=420$ MeV.
The valence quarks are realized using nonperturbatively improved Wilson-clover fermions. For those
ensembles in which the gauge and fermion fields fulfill open boundary conditions in the time direction, 
we make the correlator measurements in the bulk of the lattice where the effects of finite temporal 
extent are sufficiently damped. We distribute the source time slices evenly along the temporal dimension
for the rest of the ensembles with periodic boundary conditions. In Figure \ref{ensmap}, we show the list
of ensembles for which we obtained the results presented here. More ensembles are in our production plan 
to constrain the infinite-volume physics with good control over systematics.

The left side of Figure \ref{ensmap} is a scatter plot of all the ensembles, with the y-axis referring 
to the physical lattice extension and the x-axis gives the info on the lattice spacings. The main reason
for our choice of ensembles is to extract finite-volume spectra in multiple volumes to constrain the
scattering amplitudes more precisely. The same ensembles are also shown with the y-axis indicating the
respective pion masses on the right side of Figure \ref{ensmap}. The dotted gray line represents the
$SU(3)_f$ symmetric case, whereas the solid line at the bottom indicates the physical pion mass limit. 
As is evident from the figure, we utilize ensembles with different pion masses (equivalently different 
extents of $SU(3)_f$ symmetry breaking) to investigate the fate of $H$ dibaryon at different physical
situations. 

\begin{table}[ht]
\centering
\begin{tabular}{ccccccccc}
  \hline
  ID & $\beta$ & $N_s$ & $N_t$ & $m_{\pi}$[MeV] & $N_{\rm cfgs}$ & $N_{\rm LapH}$ & $N_{\rm tsrc}$  \\
  \hline 
  U102 & 3.40 & 24 & 128 & 350 & 4861 & 20 & 5 \\
  H102 & 3.40 & 32 & 96 & 350 & 2005 & 48 & 4 \\
  \hline 
  N200 & 3.55 & 48 & 128 & 280 & 1712 & 68 & 8 \\
  N451 & 3.46 & 48 & 128 & 280 & 1011 & 108 & 8 \\
  \hline 
  D200 & 3.55 & 64 & 128 & 200 & 2001 & 448$^*$ & 1 \\
  \hline
\end{tabular}
\caption{The details of lattice QCD ensembles referred to in this talk. $N_{\rm LapH}$ is the 
number of Laplacian eigenvectors utilized for the distillation procedure, and $N_{\rm tsrc}$ is 
the number of source time slices used. $^*$For the D200 ensemble, we utilize the stochastic LapH 
technique in which the Laplacian eigenvectors are interlaced with 16 dilution projectors and with 
full spin dilution.}\label{ensdets}
\end{table}

\textbf{Construction of correlation matrices}:
We employ the standard distillation technique to evaluate the correlation functions/matrices, except for 
the D200 ensemble. The large physical volume $V=(4.11 \mbox{ fm})^3$ of the D200 ensemble demands a large 
number of Laplacian eigenmodes $N_{LapH}$ to be used in the distillation framework. To this end,
the investigations on D200 are performed using the stochastic form of distillation technique to 
circumvent the huge computational demands due to the use of a large $N_{LapH}$. In Table
\ref{ensdets}, we present the relevant details of ensembles for which results are presented in this talk. 

\textbf{Interpolating operators}:
Throughout these calculations, we utilize only baryon-baryon interpolators in which each baryon is
separately projected to definite momentum. The general form of the momentum projected single baryon 
operators looks like 
\begin{equation}
{\cal B}_\mu(\mathbf{p},t)[q_1q_2q_3]\!=\!\sum_{\mathbf{x}} \epsilon_{abc} [q_1^{aT}(\mathbf{x},t)  
C\gamma_5 P_+q_2^b (\mathbf{x},t)] ~[ q_3^c(\mathbf{x},t)]_{\mu}~\mathrm{e}^{i\mathbf{x\cdot p}}.
\label{sB}
\end{equation}
Here $C$ is the charge conjugation operator, and $P_+=\frac{1}{2}(1+\gamma_0)$ projects the quark fields 
to positive parity. The two baryon operators are built from these single baryon interpolators using 
$\Gamma = C\gamma_5P_+$ and $\Gamma=C\gamma_iP_+$ to form the spin-zero and spin-one configurations,
respectively, as follows
\begin{equation}
[{\cal B}^{(1)}{\cal B}^{(2)}](\mathbf{p}_1,\mathbf{p}_2,t) = {\cal B}^{(1)}(\mathbf{p}_1,t) \Gamma {\cal B}^{(2)}(\mathbf{p}_2,t) .
\end{equation}

At the $SU(3)_f$ symmetric point, the flavor of a system of two octet baryons can be characterized
as belonging to the following irreducible representations (irreps), $\x{8}\otimes \x{8}= (\x{1}\oplus\x{8}\oplus\x{27})_S \oplus (\x{8}\oplus\x{10}\oplus\bar{\x{10}})_A$ with $H$ dibaryon
a scalar in $\x{1}_S$. Away from the $SU(3)_f$ symmetric point, the relevant quantum numbers are
strangeness $S=-2$ and isospin $I=0$, which has contributions from $\x{1}_S$, $\x{8}_S$, and $\x{27}_S$. 
Using the (S, I) basis for individual baryons, the three relevant scattering channels are 
$\Lambda\Lambda$, $N\Xi$, and $\Sigma\Sigma$. We systematically include an interpolator for each 
low-lying noninteracting level from all three channels. Note that unlike $\Lambda\Lambda$ and
$\Sigma\Sigma$, $N\Xi$ has nonidentical particles and thus appears in both symmetric and antisymmetric
combinations. Owing to the reduced rotational symmetry on the lattice, we ensure that two-baryon operators
transform according to the finite-volume symmetry group irreps. Combining flavor, single-baryon momenta,
and spin yields a large set of interpolating operators, for which we compute correlation matrices
$C_{ij}(t)=\langle O_i(t+t_{\rm src})O_j^{\dagger} (t_{\rm src}) \rangle$. Correlation functions for 
the single baryon operators are also computed to determine the noninteracting finite-volume spectrum. 

\begin{figure}[ht!]
\centering
 \includegraphics[height=6.0cm,width=7cm]{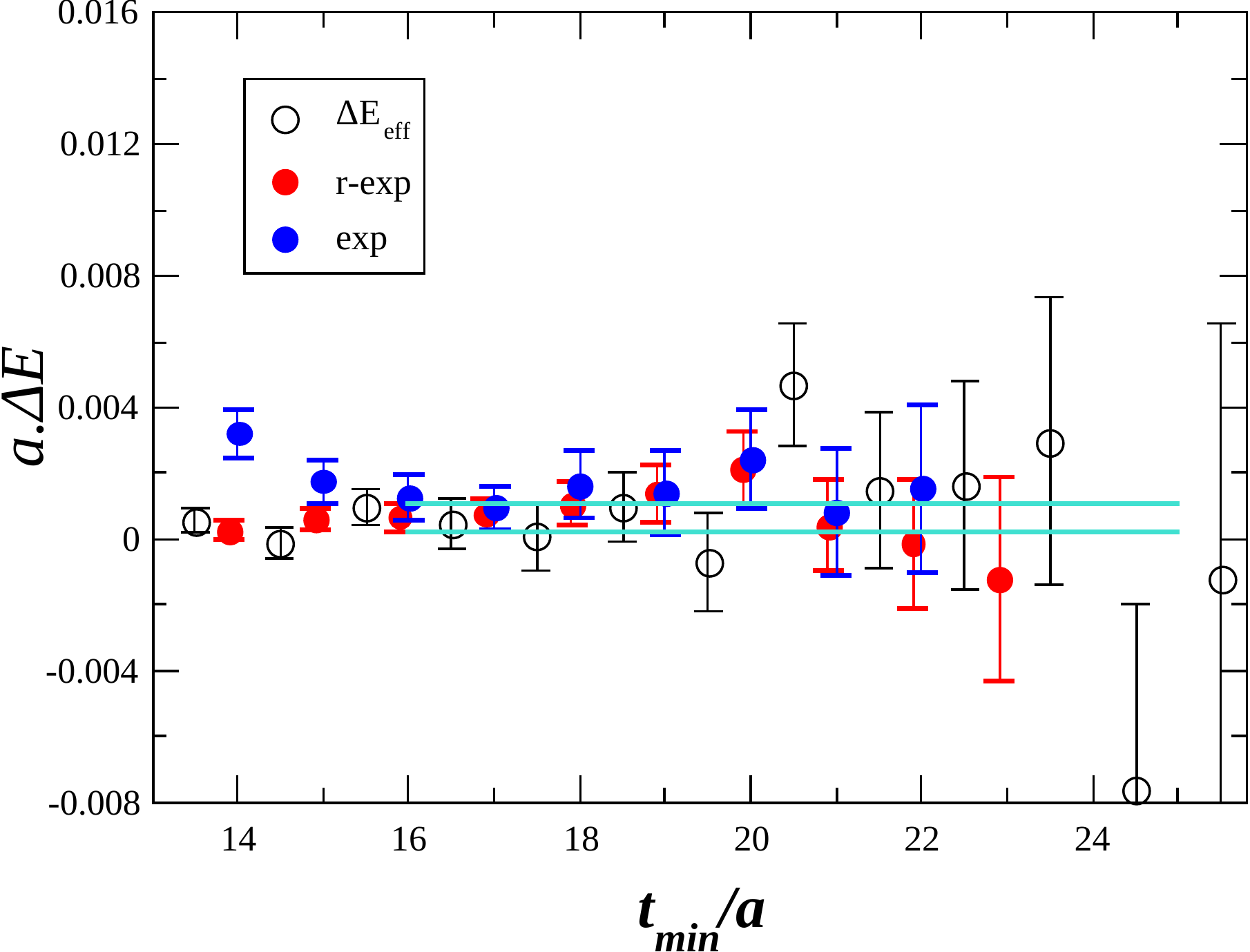}
 \caption{Comparative study of single exponential fits to $\lambda^{(n)}(t)$~[exp] and 
 $r^{(n)}(t)$~[r-exp] for the first excited state in the $A_1$ irrep of $P^2=2$ moving frame 
 in the N200 ensemble. $\Delta\mbox{E}_{\text{eff}}$ is the effective energy difference and $t_\text{min}$ refers to the boundary of the chosen fit range close to the source time slice. The cyan horizontal line
 indicates the chosen fit.}
 \label{fitquality}
\end{figure}
\textbf{Spectrum extraction}:
The finite-volume spectrum is extracted from the correlation matrices by solving the Generalized EigenValue 
Problem (GEVP) %\cite{Michael:1985ne}
\begin{equation}
C_{ij}(t)v^{(n)}_j(t,t_0) = \lambda^{(n)}(t,t_0)C_{ij}(t_0)v^{(n)}_j(t,t_0).
\end{equation}
Here the size of the correlation matrix ($m$) is as large as 28 in some of the finite-volume little group 
irreps we have considered. In the large time limit, the eigenvalue
correlators $\lambda^{(n)}(t,t_0)$ are saturated by the lightest $m$ states and can be shown to 
have an asymptotic form of $\lambda^{(n)}(t,t_0)\propto e^{-E_nt}$. An early $t_0$ is chosen such
that the noise in $C(t_0)$ does not enter the eigensolutions while also ensuring that the
extracted finite-volume spectrum is robust with its variation. The eigenvalues at sufficiently
large times are then fit with a single exponential to extract the energy spectrum.

The best fits are chosen based on a comparative study between fits to the eigenvalue correlators
$\lambda^{(n)}$ and their ratios [$r^{(n)} = \lambda^{(n)}/(C_{{\cal B}^{(1)}}C_{{\cal 
B}^{(2)}})$] with a nearby noninteracting level [${\cal B}^{(1)}{\cal B}^{(2)}$]. In Figure 
\ref{fitquality}, we present the effective energy difference ($\Delta\mbox{E}_{\text{eff}}$) 
given by $\ln(\frac{r^{(n)}(t)}{r^{(n)}(t+1)})$ along with the energy splitting estimates from 
the single exponential fits to $\lambda^{(n)}(t)$~[exp] and $r^{(n)}(t)$~[r-exp], for the first 
excited state in the $P^2=2$ moving frame on the N200 ensemble. The energy splittings from the 
fits to $\lambda^{(n)}$ are built using the energies for single hadrons determined from separate
fits to the single hadron correlators $(C_{{\cal B}^{(1)}}~\&~C_{{\cal B}^{(2)}})$. Our final choices
are generally made with the ratio fits, and such a comparative study ensures that the chosen fit
ranges are robust in terms of the ground state signal saturation.

\section{Results}\label{results}

\begin{figure}
 \centering
 \includegraphics[height=8cm,width=13.0cm]{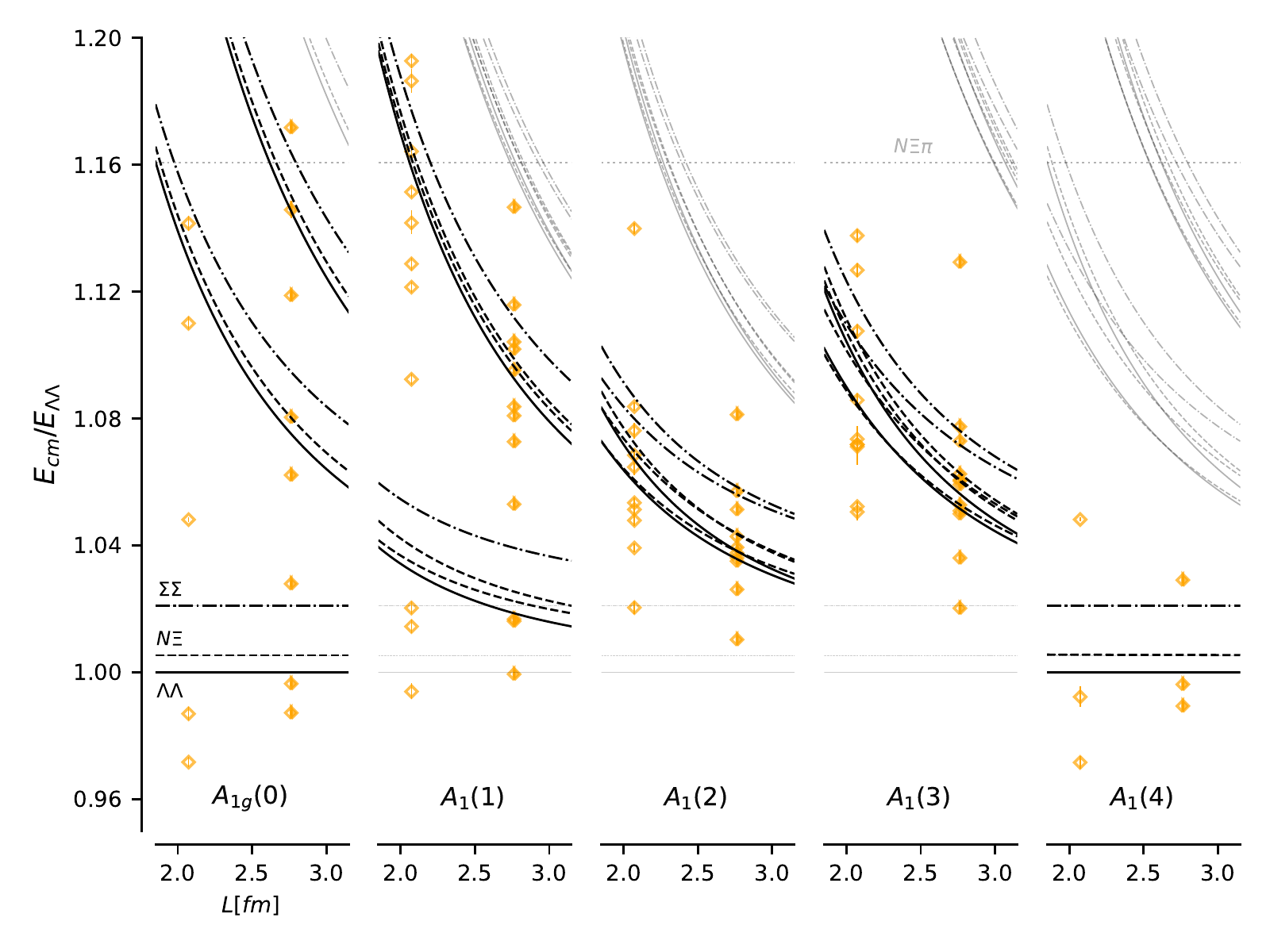}
 \caption{Energy spectrum of $I=0, S=-2$ dibaryons in the trivial finite-volume irreps ($A_1$) in 
 the ensembles with $m_{\pi}\sim 350$ MeV. Half-filled (unfilled) markers refer to the H102 (U102) ensemble.}
 \label{spectrum_350MeV}
\end{figure}

\begin{figure}
 \centering
 \includegraphics[height=8cm,width=13.0cm]{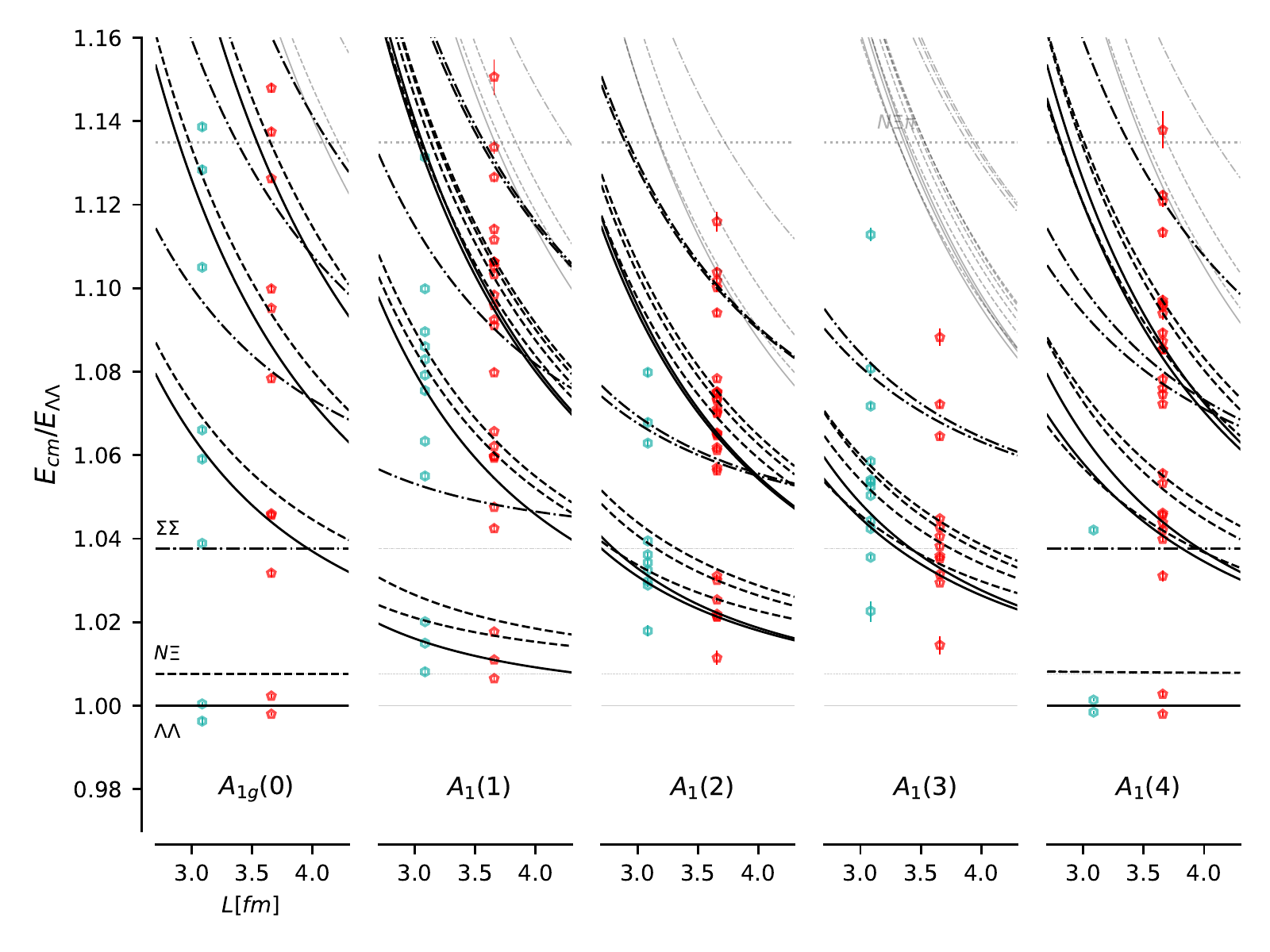}
 \caption{Same as in Figure \ref{spectrum_350MeV}, but for ensembles with $m_{\pi}\sim 280$ MeV. Red (cyan) markers refer to the N451 (N200) ensemble.}
 \label{spectrum_280MeV}
\end{figure}

\begin{figure}
 \centering
 \includegraphics[height=8cm,width=13.0cm]{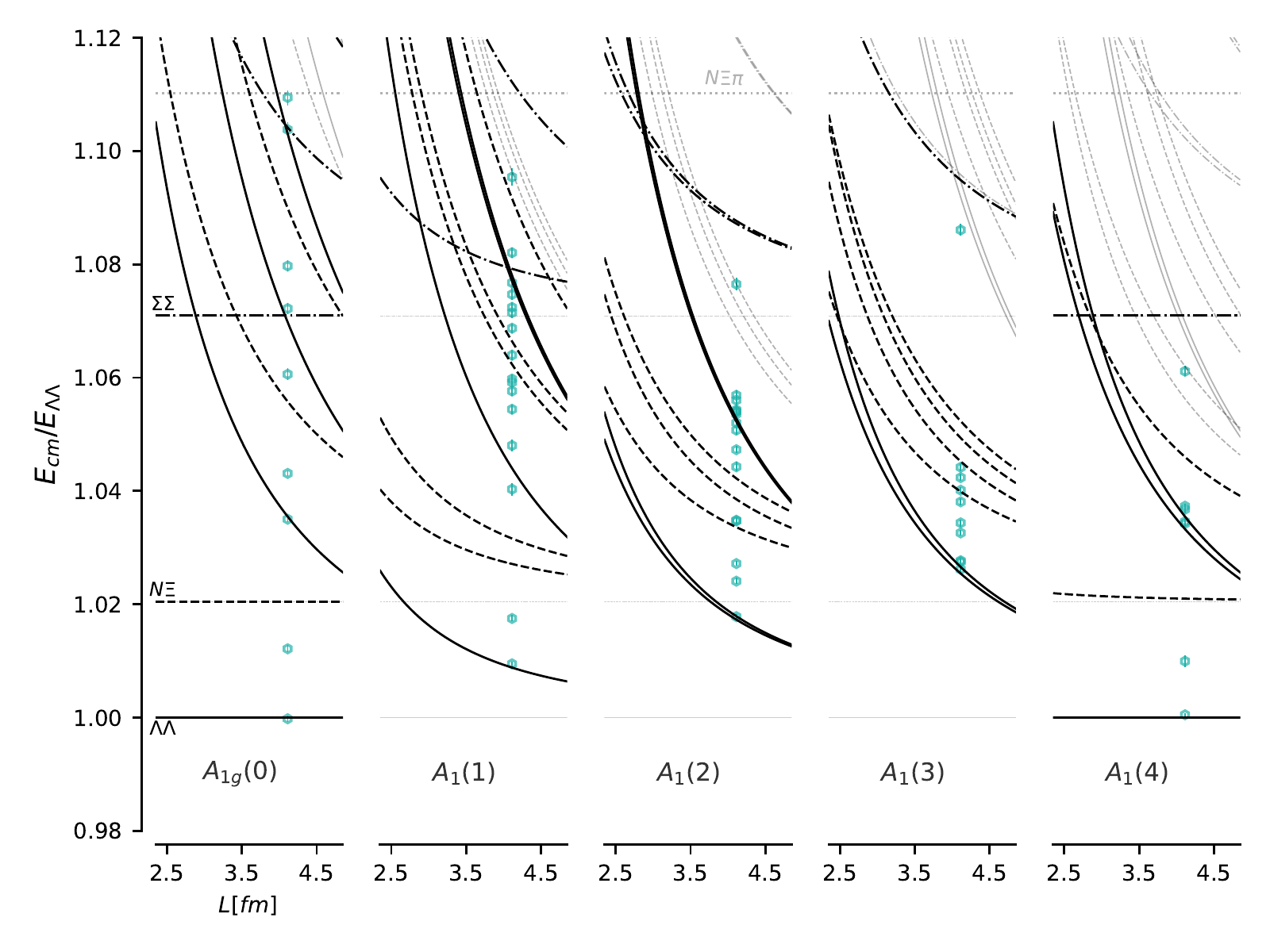}
 \caption{Same as in Figure \ref{spectrum_350MeV}, but for the D200 ensemble, which has $m_{\pi}\sim 200$ MeV.}
 \label{spectrum_200MeV}
\end{figure}

In Figures \ref{spectrum_350MeV}, \ref{spectrum_280MeV}, and \ref{spectrum_200MeV},  we present
the finite-volume energy spectrum on the five ensembles listed in the previous section. The energy
spectrum in the center-of-momentum frame is shown along the $y$-axis in units of the elastic 
threshold ($2m_{\Lambda}$). In these units, the elastic threshold always appears at the value 1. 
The $x$-axis refers to the physical lattice size in femtometers, and different panes stand for 
different finite-volume little group irreps. Upon breaking of the $SU(3)_f$ symmetry, there are 
three relevant 2-particle scattering channels ($\Lambda\Lambda$, $N\Xi$ and $\Sigma\Sigma$). The 
black and gray curves show the related noninteracting finite-volume levels. The solid curves refer
to $\Lambda\Lambda$, the dashed curves stand for $N\Xi$, and the dot-dashed are 
$\Sigma\Sigma$. The operators related to the black curves are included in the analysis, and those 
related to the gray curves are not. The lowest three-particle scattering threshold $N\Xi\pi$ is 
also shown in the figures. 

In Figure \ref{spectrum_350MeV}, we present the finite-volume energy spectrum for the ensembles 
with $m_{\pi}=350$\,MeV. Due to the proximity of the $SU(3)_f$ symmetric point, the thresholds of 
the three scattering channels are close to each other. Currently, we have results from two 
ensembles at the same lattice spacing. The energy spectrum for the $m_{\pi}=280$\,MeV ensembles is
shown in Figure\,\ref{spectrum_280MeV}. In this case, we have data at two different lattice 
spacings. For the ensemble with a larger physical volume, we have utilized a larger basis of 
baryon-baryon interpolators to extract an equally large tower of excited states across all the 
finite-volume irreps. Note that with decreasing pion mass, the extent of $SU(3)_f$ symmetry 
breaking increases. Consequently the energy splitting between the thresholds of two-baryon 
scattering channels also increases. Larger energy splittings between the scattering channels are 
evident in the finite-volume spectrum for the ensemble with $m_{\pi}=200$ MeV, which is shown in 
Figure \ref{spectrum_200MeV}. 

Following the reliable extraction of the finite-volume energy spectra, the next thing to do is 
to extract the infinite-volume physics. We follow a procedure to extract the two-particle scattering 
amplitudes from the finite-volume spectrum through the quantization condition \cite{Briceno:2014oea}
\begin{equation}
\det(K^{-1}-B) = 0,
  \label{eq:2-3}
\end{equation}
first derived by L\"uscher for elastic scattering of two spinless particles in the rest frame 
\cite{Luscher:1986pf}. With three low lying 2-baryon scattering channels  ($\Lambda\Lambda$,
$N\Xi$ and $\Sigma\Sigma$) in the broken $SU(3)_f$ symmetry scenario, one has to deal with a 
scattering matrix of dimension >3. Assuming that higher partial wave contributions do not
influence the $s$-wave scattering in the moving frames, one could work with a 3x3 scattering 
matrix. One could further simplify the problem by assuming that effects from the
$\Sigma\Sigma$ channel are negligible. However, the applicability of this assumption is limited to lighter 
$m_{\pi}$ scenarios, owing to the greater extent of $SU(3)_f$ symmetry breaking. $N\Xi$, being a 
channel with nonidentical particles, allows mixing of spin sectors ($S=0$ and $S=1$), which in 
turn allows for physical mixing of higher partial waves unlike in the $SU(3)_f$ symmetric case. 
Note that in moving frames, the first higher partial wave that can contribute to the finite-volume
spectra is the $p$-wave. Relaxing the assumptions on neglecting higher partial wave effects 
complicates the problem of quantization further due to an enlarged scattering matrix. 

The extracted finite-volume energy spectra are very dense, and several energy levels are nearly 
degenerate. Standard procedures such as minimizing the Determinant Residual 
\cite{Morningstar:2017spu} or a $\chi^2$ defined from the extracted finite-volume energy spectrum
and the reconstructed energy spectrum from the zeros of the quantization determinant 
\cite{Dudek:2014qha} are reaching their limits with such a dense spectrum. Currently, we are 
working on realizing a newer analysis procedure utilizing the eigenvalue decomposition of the 
quantization matrix \cite{Woss:2020cmp}, which we believe is the way to go forward with a 
complicated system such as this\footnote{We utilize the {\it TwoHadronsInBox} package  to realize the
quantization condition \cite{Morningstar:2017spu}.}. In addition to the fact that this is a system 
involving multi-channel scattering, we also need to be cautious about various systematic uncertainties 
that could be crucial. Our experience from the studies made at the $SU(3)_f$ symmetric point suggests 
that there could be large discretization effects \cite{Green:2021qol}. Furthermore, the experimental 
bounds and the lessons from our studies at the $SU(3)_f$ symmetric point suggest that the continuum 
binding energy of $H$ dibaryon, if it exists, could be very small. There is no reason to expect a 
different scenario in the $SU(3)_f$ broken situation, at least for the chosen discretization. These
observations call for lattice calculations with good control over the systematic uncertainties. To this
end, we plan to extend our investigations to several ensembles over a wide range of lattice spacings and 
volumes. 

\section{Summary}
We have reported preliminary results for $H$~dibaryon spectroscopy away from the $SU(3)_f$ symmetric point,
obtained by applying the distillation framework on a set of ensembles with $N_f=2+1$ flavors of 
${\cal{O}}(a)$-improved Wilson quarks, generated by CLS. We are able to resolve a dense spectrum of 
finite-volume energy levels at several values of the pion mass. Current efforts focus on the extraction of 
infinite-volume scattering amplitudes by applying the finite-volume quantization condition. We will also 
extend our analysis to dibaryon systems other than the $H$~dibaryon, for which the correlator data have 
already been computed.

\acknowledgments
Calculations for this project used resources on the supercomputers
JUQUEEN~\cite{juqueen}, JURECA~\cite{jureca}, and JUWELS~\cite{juwels}
at Jülich Supercomputing Centre (JSC) and Frontera at the Texas Advanced Computing Center (TACC). The authors gratefully
acknowledge the support of the John von Neumann Institute for
Computing and Gauss Centre for Supercomputing e.V.\
(\url{http://www.gauss-centre.eu}) for project HMZ21.
This research is partly supported by Deutsche Forschungsgemeinschaft (DFG, German Research Foundation) through the Collaborative Research Center SFB 1044 ``The low-energy frontier of the Standard Model'' and  the Cluster of Excellence ``Precision Physics, Fundamental Interactions and Structure of Matter'' (PRISMA$^+$, EXC 2118/1) funded by DFG within the German Excellence Strategy (Project ID 39083149).
ADH is supported by the U.S. Department of Energy, Office of Science, Office of Nuclear Physics through
the Contract No. DE-SC0012704 and within the framework of Scientific Discovery through
Advance Computing (SciDAC) award ``Computing the Properties of Matter with Leadership Computing Resources.''
The work of BH was supported by an LBNL LDRD Award.
CJM acknowledges support from the U.S.~NSF under award PHY-1913158.
We are grateful to our colleagues within the CLS initiative for sharing ensembles. PM is grateful to  Andr\'e~Walker-Loud for careful reading of the manuscript. 

\bibliographystyle{JHEP}
\bibliography{references}

\providecommand{\href}[2]{#2}\begingroup\raggedright\begin{thebibliography}{10}

\bibitem{Jaffe:1976yi}
R.~L. Jaffe, \href{https://doi.org/10.1103/PhysRevLett.38.617,
  10.1103/PhysRevLett.38.195}{\emph{Phys. Rev. Lett.} {\bfseries 38} (1977)
  195}.

\bibitem{Takahashi:2001nm}
H.~Takahashi et~al.,
  \href{https://doi.org/10.1103/PhysRevLett.87.212502}{\emph{Phys. Rev. Lett.}
  {\bfseries 87} (2001) 212502}.

\bibitem{ALICE:2019eol}
{\scshape ALICE} collaboration, S.~Acharya et~al., ,
  \href{https://doi.org/10.1016/j.physletb.2019.134822}{\emph{Phys. Lett. B}
  {\bfseries 797} (2019) 134822}
  [\href{https://arxiv.org/abs/1905.07209}{{\ttfamily 1905.07209}}].

\bibitem{Mackenzie:1985vv}
P.~B. Mackenzie and H.~B. Thacker,
  \href{https://doi.org/10.1103/PhysRevLett.55.2539}{\emph{Phys. Rev. Lett.}
  {\bfseries 55} (1985) 2539}.

\bibitem{Inoue:2010es}
{\scshape HAL QCD} collaboration, T.~Inoue, N.~Ishii, S.~Aoki, T.~Doi,
  T.~Hatsuda, Y.~Ikeda et~al., ,
  \href{https://doi.org/10.1103/PhysRevLett.106.162002}{\emph{Phys. Rev. Lett.}
  {\bfseries 106} (2011) 162002}
  [\href{https://arxiv.org/abs/1012.5928}{{\ttfamily 1012.5928}}].

\bibitem{NPLQCD:2010ocs}
{\scshape NPLQCD} collaboration, S.~R. Beane et~al., ,
  \href{https://doi.org/10.1103/PhysRevLett.106.162001}{\emph{Phys. Rev. Lett.}
  {\bfseries 106} (2011) 162001}
  [\href{https://arxiv.org/abs/1012.3812}{{\ttfamily 1012.3812}}].

\bibitem{NPLQCD:2012mex}
{\scshape NPLQCD} collaboration, S.~R. Beane, E.~Chang, S.~D. Cohen,
  W.~Detmold, H.~W. Lin, T.~C. Luu et~al., ,
  \href{https://doi.org/10.1103/PhysRevD.87.034506}{\emph{Phys. Rev. D}
  {\bfseries 87} (2013) 034506}
  [\href{https://arxiv.org/abs/1206.5219}{{\ttfamily 1206.5219}}].

\bibitem{Francis:2018qch}
A.~Francis, J.~R. Green, P.~M. Junnarkar, C.~Miao, T.~D. Rae and H.~Wittig,
  \href{https://arxiv.org/abs/1805.03966}{{\ttfamily 1805.03966}}.

\bibitem{Green:2021qol}
J.~R. Green, A.~D. Hanlon, P.~M. Junnarkar and H.~Wittig,
  \href{https://arxiv.org/abs/2103.01054}{{\ttfamily 2103.01054}}.

\bibitem{Jeremy:2021mai}
J.~R. Green et~al., {\emph{PoS} {\bfseries LATTICE2021} (2021) 294}.

\bibitem{Briceno:2014oea}
R.~A. Briceño, \href{https://doi.org/10.1103/PhysRevD.89.074507}{\emph{Phys.
  Rev. D} {\bfseries 89} (2014) 074507}
  [\href{https://arxiv.org/abs/1401.3312}{{\ttfamily 1401.3312}}].

\bibitem{Luscher:1986pf}
M.~Lüscher, \href{https://doi.org/10.1007/BF01211097}{\emph{Commun. Math.
  Phys.} {\bfseries 105} (1986) 153}.

\bibitem{Morningstar:2017spu}
C.~Morningstar, J.~Bulava, B.~Singha, R.~Brett, J.~Fallica, A.~Hanlon et~al.,
  \href{https://doi.org/10.1016/j.nuclphysb.2017.09.014}{\emph{Nucl. Phys. B}
  {\bfseries 924} (2017) 477}
  [\href{https://arxiv.org/abs/1707.05817}{{\ttfamily 1707.05817}}].

\bibitem{Dudek:2014qha}
{\scshape HS} collaboration, J.~J. Dudek et~al., ,
  \href{https://doi.org/10.1103/PhysRevLett.113.182001}{\emph{Phys. Rev. Lett.}
  {\bfseries 113} (2014) 182001}
  [\href{https://arxiv.org/abs/1406.4158}{{\ttfamily 1406.4158}}].

\bibitem{Woss:2020cmp}
{\scshape Hadron Spectrum} collaboration, A.~J. Woss, D.~J. Wilson and J.~J.
  Dudek, , \href{https://doi.org/10.1103/PhysRevD.101.114505}{\emph{Phys. Rev.
  D} {\bfseries 101} (2020) 114505}
  [\href{https://arxiv.org/abs/2001.08474}{{\ttfamily 2001.08474}}].

\bibitem{juqueen}
{J{\"u}lich Supercomputing Centre},
  \href{https://doi.org/10.17815/jlsrf-1-18}{\emph{J. Large-Scale Res. Facil.}
  {\bfseries 1} (2015) A1}.

\bibitem{jureca}
{J{\"u}lich Supercomputing Centre},
  \href{https://doi.org/10.17815/jlsrf-4-121-1}{\emph{J. Large-Scale Res.
  Facil.} {\bfseries 4} (2018) A132}.

\bibitem{juwels}
{J{\"u}lich Supercomputing Centre},
  \href{https://doi.org/10.17815/jlsrf-5-171}{\emph{J. Large-Scale Res. Facil.}
  {\bfseries 5} (2019) A135}.

\end{thebibliography}\endgroup

\end{document}